\documentclass[english]{article}
\usepackage[T1]{fontenc}
\usepackage{textcomp}
\usepackage{geometry}
\usepackage{babel}
\usepackage{varwidth}
\usepackage{amsmath}
\usepackage{stackrel}
\usepackage{graphicx}
\usepackage[pdfusetitle,
 bookmarks=true,bookmarksnumbered=false,bookmarksopen=false,
 breaklinks=false,pdfborder={0 0 1},backref=false,colorlinks=false]
 {hyperref}

\makeatletter

\providecommand{\tabularnewline}{\\}

\newenvironment{cellvarwidth}[1][t]
    {\begin{varwidth}[#1]{\linewidth}}
    {\@finalstrut\@arstrutbox\end{varwidth}}

\usepackage{authblk}
\usepackage{csquotes}
\usepackage{braket}
\usepackage{algorithm}
\usepackage{algpseudocode}
\usepackage{sidecap}

\ifdefined\showcaptionsetup

 \PassOptionsToPackage{caption=false}{subfig}
\fi
\usepackage{subfig}
\makeatother

\usepackage[numbers]{natbib}
\usepackage{doi}

\begin{document}
\title{Gate Optimization of NEQR Quantum Circuits via PPRM Transformation}
\author[1]{Shahab Iranmanesh}
\author[2]{Hossein Aghababa}
\author[3]{Kazim Fouladi}
\affil[1]{School of Electrical and Computer Engineering, College of Engineering, University of Tehran, Tehran, Iran \\ \href{mailto:shahab.iranmanesh@ut.ac.ir}{shahab.iranmanesh@ut.ac.ir}}
\affil[2]{Department of Engineering, Loyola University Maryland, Maryland, USA\\Founder of Quantum Computation and Communication Laboratory, University of Tehran, Tehran, Iran}
\affil[3]{Faculty of Engineering, College of Farabi, University of Tehran, Tehran, Iran}
\date{}
\maketitle
\begin{abstract}
Quantum image representation (QIR) is a key challenge in quantum image
processing (QIP) due to the large number of pixels in images, which
increases the need for quantum gates and qubits. However, current
quantum systems face limitations in run-time complexity and available
qubits. This work aims to compress the quantum circuits of the Novel
Enhanced Quantum Representation (NEQR) scheme by transforming their
Exclusive-Or Sum-of-Products (ESOP) expressions into Positive Polarity
Reed-Muller (PPRM) equivalents without adding ancillary qubits. Two
cases of run-time complexity---exponential and linear--- are considered
for NEQR circuits with $m$ controlling qubits $(m\rightarrow\infty)$,
depending on the decomposition of multi-controlled NOT gates. Using
nonlinear regression, the proposed transformation is estimated to
reduce the exponential complexity from $O(2^{m})$ to $O(1.5^{m})$,
with a compression ratio approaching 100\%. For linear complexity,
the transformation is estimated to halve the run-time, with a compression ratio approaching 52\%. Tests on six $256\times256$ images show average reductions of 105.5 times for exponential cases and 2.4 times for linear cases, with average compression ratios of 99.05\% and 58.91\%, respectively.

\vspace{1em} \noindent \textbf{Keywords:} Quantum Image Representation, NEQR, Quantum Image Compression, Multi Controlled-NOT gates, Exclusive-Or Sum-of-Products (ESOP), Optimization, Minimization, Positive Polarity Reed-Muller (PPRM), Nonlinear Regression
\end{abstract}

\section{Introduction}

Limitation is a concept intertwined with today\textquoteright s quantum
computers, preventing them from becoming general-purpose machines
and substituting digital computers. The noisy intermediate-scale quantum (NISQ) era has been marked by a limited number of qubits, a noisy and error-prone environment, and quantum decoherence, all of which restrict the design of quantum circuits in terms of execution time. Consequently, optimizing the run-time complexity of quantum circuits by reducing the number of quantum gates and changing their types is both significant and necessary.

Quantum image representation (QIR), which involves encoding digital
images into quantum circuits, is a subfield of quantum image processing
(QIP) that faces run-time challenges due to the large and increasing
number of pixels in modern images and videos. Various QIR models have
been proposed, including Qubit Lattice \cite{rep1}, Real Ket \cite{rep2},
entangled images \cite{rep8}, and a Flexible Representation of Quantum
Images (FRQI), which encodes pixel grayscale values into a grayscale
qubit by mapping each pixel value to an angle \cite{rep3}. Other
models include an RGB Multi-Channel Representation for Quantum Images
(MCQI) \cite{rep9} and a Novel Enhanced Quantum Representation (NEQR)
\cite{neqr}, which is similar to classical image representation and
facilitates classically inspired arithmetic operations, with the potential
for processing quantum operations on all pixels simultaneously due
to the quantum mechanics superposition phenomenon. Among these models,
FRQI and NEQR are more commonly used, and some representations have
been developed based on these two, such as an improved FRQI model
(FRQCI) \cite{rep6}, Improved NEQR (INEQR) \cite{rep4}, a Generalized
Quantum Image Representation (GQIR) \cite{rep7}, and a Generalized
model of NEQR (GNEQR) \cite{rep5}.

To decrease the run-time complexities of QIR schemes, researchers
have proposed various optimizations and compressions, referred to
as quantum image compressions (QICs). QIC aims to reduce the quantum
resources required for QIR by compressing a digital image before constructing
its quantum circuit, lowering the number of quantum gates, and simplifying
quantum gates \cite{qic}. A few QICs have been proposed, such as
a hybrid quantum vector quantization encoding algorithm \cite{compress4},
using the Quantum Discrete Cosine Transform (QDCT) \cite{compress5},
image compression based on the Functional Sized Population Quantum
Evolutionary Algorithm (FSQEA) \cite{compress6}, an algorithm using
Quantum Backpropagation (QBP) \cite{compress7}, employing Grover\textquoteright s
quantum search algorithm \cite{compress8}, the DCT-GQIR method which
uses Direct Cosine Transform preparation and GQIR encoding inspired
by JPEG image compression \cite{compress1}, applying compression
to quantum RGB images using the quantum Haar wavelet transform (HQWT)
and iterative quantum Fibonacci transform (IQFT) \cite{compress9},
a quantum version of an autoencoder based on parameterized quantum
circuits \cite{compress10}, and an image compression and reconstruction
algorithm leveraging the quantum network (QN) and the gradient descent
algorithm \cite{compress11}. Other QICs focus on minimizing QIR or
both image compression and QIR minimization. Enhanced FRQI (EFRQI)
minimizes GQIR representation using auxiliary qubits \cite{compres2},
while the DCT-EFRQI scheme compresses images and minimizes their QIRs
\cite{compres3}. This present work aims to optimize NEQR quantum
circuits without any compression before the quantum representation
stage.

An NEQR circuit consists of a collection of quantum logic multi-controlled
NOT (MCNOT) gates, mapping the circuit\textquoteright s output to
Exclusive-Or Sum-of-Products (ESOP) expansions. Some works suggest
using the Espresso algorithm \cite{espresso} for minimizing the ESOP
expressions resulting from an NEQR quantum circuit \cite{neqr}\cite{rep10}\cite{rep11}.
However, the Espresso algorithm is best suited for minimizing Boolean
functions in the form of Sum-of-Products (SOP). To minimize the ESOP
expressions resulting from an NEQR circuit using Espresso, an additional
process is required to convert the minimized result into ESOP form.
Many methods have been introduced to optimize ESOP expressions directly,
which are either exact or heuristic, including a nonlinear integer
programming approach \cite{esop1}, a parallel algorithm \cite{esop2},
using ordered Kronecker functional decision diagrams \cite{esop3},
fast heuristic minimization \cite{esop4}, and Karnaugh Maps \cite{karno}\cite{khodam}.

In this work, transforming ESOP expressions, resulting from implementing 
NEQR circuits, to Positive Polarity Reed-Muller (PPRM) expressions
without further optimization is investigated. Furthermore, to measure
and compare the run-time complexities of non-optimized NEQR circuits
and their optimized versions, the concept of quantum cost (QC) is
defined based on the number of primitive quantum gates, such as the
CNOT gate, whose QC is considered 1. Two cases of compression rate
and QC optimization rate are explored for images optimized by the
PPRM transformation and are estimated using nonlinear regression to
derive mathematical functions that verify their growth as the number
of pixels increases. To promote transparency, the source code is available on GitHub at \url{https://github.com/shahab-iranmanesh/PPRM_Optimization_of_NEQR}.

The rest of this report is organized as follows: Section \ref{sec:Background}
discusses two kinds of MCNOT gates, NEQR representations, and PPRM
expressions. Section \ref{sec:PPRM-transform} describes two algorithms
for transforming ESOP expressions into PPRM ones, which are used in
Section \ref{sec:Simulation} to explore the optimization and compression
rates for real and random images. Finally, Section \ref{sec:Conclusion}
offers some concluding remarks.

\section{Background\protect\label{sec:Background}}

\subsection{Multi-Controlled NOT (MCNOT) gate}

\begin{figure}[t]
\begin{centering}
\includegraphics[scale=0.5]{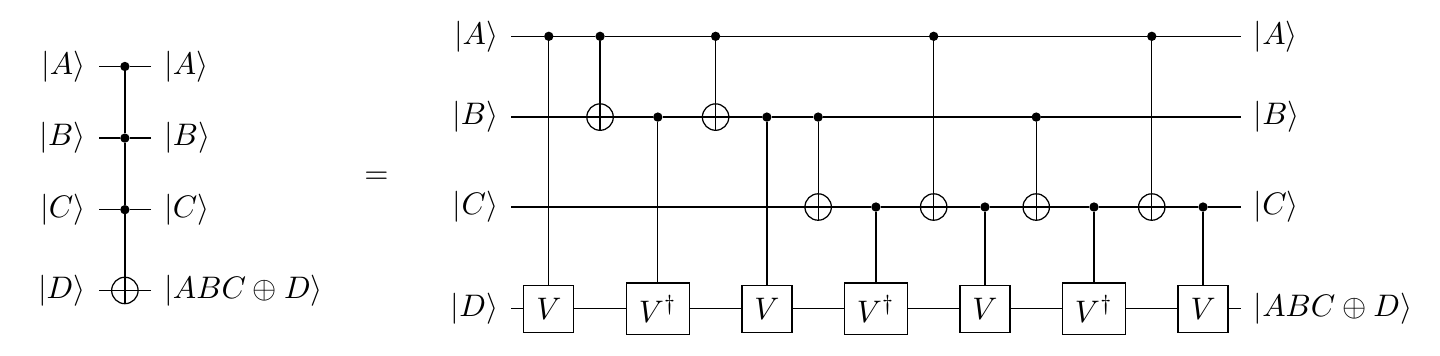}
\par\end{centering}
\caption{{\small\protect\label{fig:mcnot} Quantum circuit of the $C^{3}(X)$
gate without ancillary qubits \cite{barenco}}}
\end{figure}

\begin{figure}[t]
\begin{centering}
\includegraphics[scale=0.7]{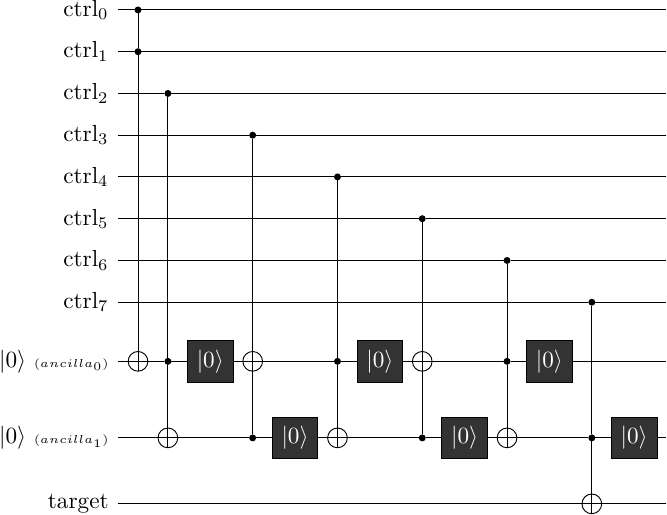}
\par\end{centering}
\caption{{\small\protect\label{fig:mcnot-r} Quantum circuit of the proposed
$C_{R}^{8}(X)$ gate using two ancillary qubits}}
\end{figure}

The MCNOT gate, also referred to as $C^{m}(X)$ in this paper, where
$X=\begin{pmatrix}0 & 1\\
1 & 0
\end{pmatrix}$, is a controlled Pauli-X gate with $m\geq2$ control qubits and one
target qubit. When all its control qubits are in the state $\ket{1}$,
the target qubit is flipped; otherwise, the target qubit retains its
initial state. Various methods have been proposed to implement the
$C^{m}(X)$ gate in quantum circuits, as it is not a universal basis
gate for quantum systems, unlike the CNOT gate. In this paper, two
designs of the $C^{m}(X)$ gate are considered for simulation and
analysis. One design, shown in Figure~\ref{fig:mcnot} and described
in \cite{barenco}, does not use any additional ancillary qubits and
has an exponential quantum cost (QC) of $3\times2^{m}-4$, as calculated
in \cite{khodam}. The other design, proposed in \cite{khodam} and
shown in Figure~\ref{fig:mcnot-r}, is called the MCNOT with Reset
(MCNOT-R or $C_{R}^{m}(X)$), which utilizes a $reset$ operation
(represented by black boxes with $\ket{0}$) to reuse its two ancillary
qubits and has a QC of $19m-32$. 

Applying each $C^{m}(X)$ or $C_{R}^{m}(X)$ gate with control qubits
$\left\{ ctrl_{i}\right\} _{i=0}^{m}$ on a target qubit $t$, results
in a state $\ket{t}=\ket{\stackrel[i=0]{m}{\prod}ctrl_{i}\:\oplus\:t}$,
where $\stackrel[i=0]{m}{\prod}ctrl_{i}$ is the AND of the Boolean
variables $\left\{ ctrl_{i}\right\} _{i=0}^{m}$. For example, if
the initial state of the target qubit is set to $\ket{0}$, applying
five MCNOT gates results in the target qubit being in a state equivalent
to the XOR of five product terms, each of which is the product of
the control qubits' states of each MCNOT gate. This type of expression
is called an Exclusive-Or Sum-of-Products (ESOP), as seen in the NEQR
quantum circuits described in the next subsection.

\subsection{NEQR}

\begin{figure}[t]
\begin{centering}
\includegraphics[scale=0.9]{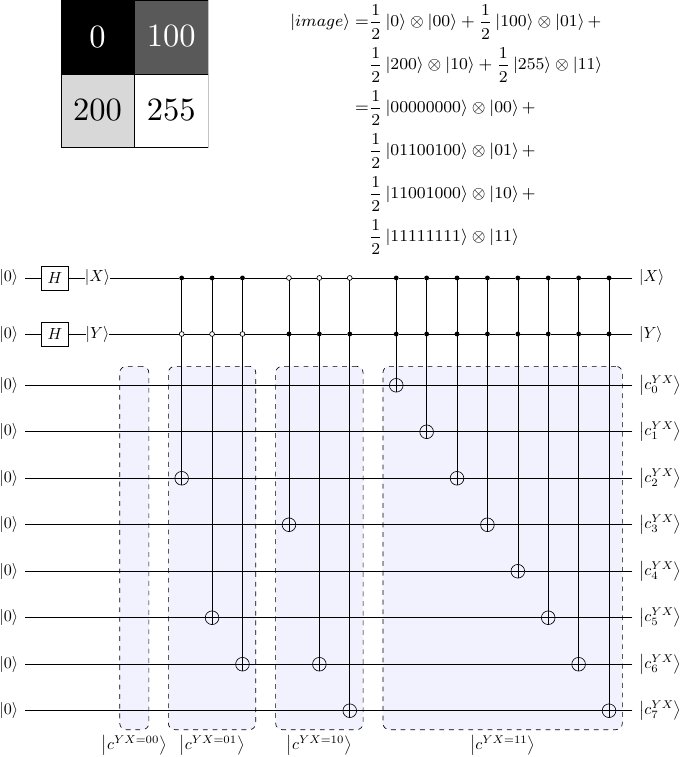}
\par\end{centering}
\caption{{\small\protect\label{fig:neqr}NEQR representation and quantum circuit
of a $2\times2$ grayscale image}}
\end{figure}

A Novel Enhanced Quantum Representation (NEQR) model \cite{neqr}
was proposed by Zhang et al. to represent a digital grayscale image
using a quantum circuit with two groups of entangled qubits: one group
for the grayscale values of the pixels and another for the coordinates
that provide positional information for all the pixels simultaneously.
A $2^{n}\times2^{n}$digital image with grayscale values in the range
$[0,2^{q}-1]$ is represented by a normalized superposition of tensor
products of quantum states, as expressed by Equation~\ref{eq:neqr_expres},
where $\ket{C_{i}^{YX}}$ represents the $i^{th}$ qubit of the $q$-qubit
grayscale value associated with the pixel at vertical coordinate $Y$
and horizontal coordinate $X$.
\begin{quote}
\begin{equation}
\ket{image}=\frac{1}{2^{n}}\ \sum_{Y=0}^{2^{n}-1}\ \sum_{X=0}^{2^{n}-1}\ \bigotimes_{i=0}^{q-1}\ket{C_{i}^{YX}}\ket{Y}\ket{X}\label{eq:neqr_expres}
\end{equation}
\end{quote}
\noindent When a grayscale bit of an image at a specified pixel is
1, the corresponding qubit must be flipped from its initial value
$\ket{0}$ to $\ket{1}$. Figure~\ref{fig:neqr} shows an example
of a $2\times2$ grayscale image, its NEQR expression, and the corresponding
quantum circuit. As illustrated, for each grayscale qubit $\ket{C_{i}^{YX}}$,
several $C^{2}(X)$ (Toffoli) gates are applied under the control
of $Y$ and $X$ qubits. Each of these control qubits is placed into
an equal superposition of $\ket{0}$ and $\ket{1}$ by a Hadamard
gate $H=\frac{1}{\sqrt{2}}\begin{pmatrix}1 & 1\\
1 & -1
\end{pmatrix}$ before controlling the Toffoli gates. For example, the grayscale
value of $\ket{C_{0}^{YX}}$ is $\ket{1}$ only when $\ket{YX}=\ket{11}$.
In this case, a single Toffoli gate is required, controlled by $Y$
and $X$ qubits in their positive state. Thus, if the superposition
of the coordinate qubits is measured and the result is 11, the value
of $\ket{C_{0}^{YX}}$ is expected to be 1 after measurement.

The larger an image is, the more coordinate qubits are required, necessitating
the use of MCNOT gates with more than two control qubits. For an NEQR
circuit representing a $2^{n}\times2^{n}$ image, MCNOT gates with
$2n$ control qubits ($C^{2n}(X)$) are needed. Consequently, such
quantum circuits can have a high run-time complexity, with either
exponential or polynomial QC, making them unsuitable for the coherence
times of current quantum systems. Therefore, optimizing NEQR circuits
is essential.

\subsection{Positive Polarity Reed-Muller (PPRM) Expressions}

ESOP expressions, also known as Reed-Muller (RM) expressions, can
be optimized using various proposed methods. As an example of a two-input
expression, consider the $\ket{C_{6}^{YX}}$ qubit in Figure~\ref{fig:neqr},
which has three Toffoli gates applied to it. This results in the expression
$0\oplus\overline{Y}X\oplus Y\overline{X}\oplus YX$. To optimize
the gates applied to $\ket{C_{6}^{YX}}$, the ESOP expression $0\oplus\overline{Y}X\oplus Y\overline{X}\oplus YX$
can be simplified to $\overline{Y}X\oplus Y$, as calculated in Equation~\ref{eq:esop}.
However, for a large number of inputs, manual calculation or other
similar methods, such as Karnaugh maps, are impractical and do not
guarantee a globally optimal solution.

\begin{equation}
0\oplus\overline{Y}X\oplus Y\overline{X}\oplus YX=0\oplus\overline{Y}X\oplus Y(\overline{X}\oplus X)=0\oplus\overline{Y}X\oplus Y=\overline{Y}X\oplus Y\label{eq:esop}
\end{equation}

\begin{SCfigure}[2.0]
  \centering
  \includegraphics[width=0.3\textwidth]{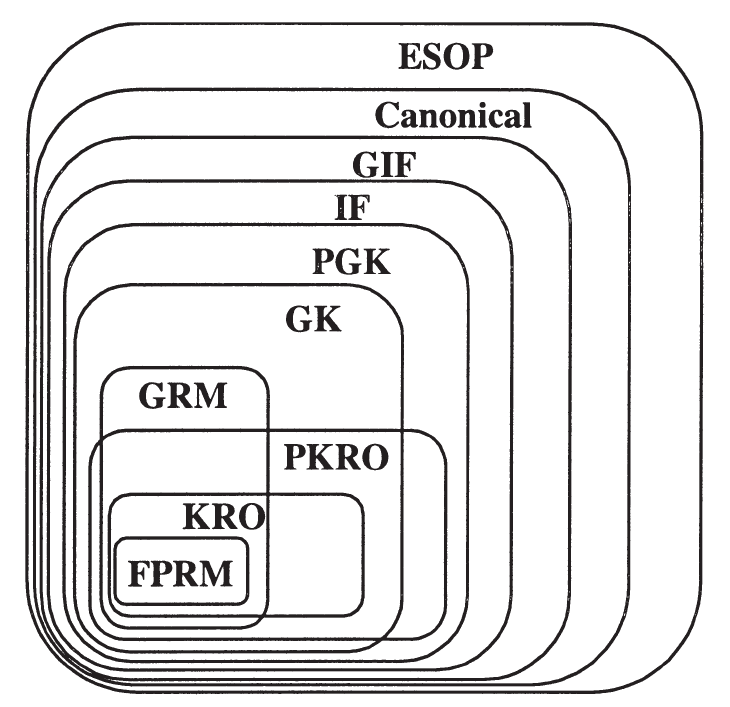}
  \caption{Set-theoretic relationships between different families of ESOP canonical forms \cite{family}}
\label{fig:family}
\end{SCfigure}

ESOP expressions have canonical subgroup families, as illustrated
in Figure~\ref{fig:family} \cite{family}. In this figure, the Fixed
Polarity Reed-Muller (FPRM) form is the smallest subgroup compared
to other subgroups, with product terms fixed in terms of polarity.
The Positive Polarity Reed-Muller (PPRM) is a form of FPRM with positive
polarity, where product terms are composed of positive and non-complementary
literals, as shown in Equation~\ref{eq:pprm}. Since the PPRM form
is the smallest subgroup among these families, it is expected that
NEQR circuits can be simplified by converting their Boolean functions
to PPRM form.

\begin{equation}
f(x_{n-1},x_{n-2},\ldots,x_{0})=\stackrel[k=0]{2^{n}-1}{\sum}a_{k}m_{k};\quad a_{k}\in\left\{ 0,1\right\} ,\quad m_{k}:minterm\label{eq:minterms}
\end{equation}
\begin{equation}
f(x_{n-1},x_{n-2},\ldots,x_{0})=b_{0}\oplus b_{1}x_{0}\oplus b_{2}x_{1}\oplus b_{3}x_{1}x_{0}\oplus\ldots\oplus b_{2^{(n-1)}}x_{n-1}x_{n-2}\ldots x_{0};\quad b_{j}\in\left\{ 0,1\right\} \label{eq:pprm}
\end{equation}

\section{PPRM transform\protect\label{sec:PPRM-transform}}

An algorithmic and precise method to obtain the PPRM form of a Boolean
expression is described as follows \cite{pprm_calc1}. A matrix named
$R(n)$ is produced by the $n$-times Kronecker product of $R(1)$
with itself, as shown in Equation~\ref{eq:R}. Then, $X(n)$ is defined
as in Equation~\ref{eq:X}, where $x_{i}$'s are the inputs of the
Boolean function $f$, as shown in Equation~\ref{eq:minterms}.

\begin{equation}
R(1)=\begin{pmatrix}1 & 0\\
1 & 1
\end{pmatrix};\quad R(n)=\stackrel[i=0]{n-1}{\bigotimes}R(1)\label{eq:R}
\end{equation}
\begin{align}
X(n)=\stackrel[i=0]{n-1}{\bigotimes}[1\quad x_{i}]=[1\quad x_{n-1}]\otimes[1\quad x_{n-2}]\otimes\ldots\otimes[1\quad x_{0}]\nonumber \\
=\left[1\quad x_{0}\quad x_{1}\quad x_{1}x_{0}\quad x_{2}\quad x_{2}x_{0}\quad x_{2}x_{1}\quad x_{2}x_{1}x_{0}\ldots\right]\label{eq:X}
\end{align}
After calculating the $2^{n}$ vector $\overrightarrow{b}$, which
includes the $b_{j}$ coefficients used in Equation~\ref{eq:pprm},
the PPRM form of $f$ can be obtained by the $*$ operation between
$X(n)$ and $\overrightarrow{b}$, which multiplies each corresponding
element and XORs them together, as shown in Equation~\ref{eq:Xb}.
To calculate $\overrightarrow{b}$, the $*$ operation is applied
between $R(n)$ and the vector $\overrightarrow{a}$, which includes
the $a_{k}$ coefficients used in Equation~\ref{eq:minterms}.

\begin{eqnarray}
\overrightarrow{b} & = & R(n)*\overrightarrow{a}\nonumber \\
f & = & X(n)*\overrightarrow{b}=\left[1\quad x_{0}\quad x_{1}\quad x_{1}x_{0}\quad\ldots\right]*\begin{pmatrix}b_{0}\\
b_{1}\\
b_{2}\\
b_{3}\\
\vdots
\end{pmatrix}=b_{0}\oplus b_{1}x_{0}\oplus b_{2}x_{1}\oplus b_{3}x_{1}x_{0}\oplus\ldots\label{eq:Xb}
\end{eqnarray}

Bakoev et al. proposed another algorithm in \cite{bakoev} to calculate
$\overrightarrow{b}$ with a dimension of $2^{n}$ more quickly than
using $R(n)$, consuming considerably less memory. As described in
Algorithm~\ref{alg:binary_pprm}, the input vector $\overrightarrow{a}$
is divided into blocks of size $2^{k}$ $(k=0,1,\ldots,n)$ by using
a dynamically changing $mask$ array, as shown in lines~\ref{line:mask1}
and~\ref{line:mask2}. In each iteration, a bitwise XOR operation
is performed between the corresponding odd and even blocks. The result
of this XOR operation is stored in the even blocks, while the odd
blocks remain unchanged. This process repeats, with the block size
doubling at each step. \begin{algorithm}
\caption{Binary PPRM Algorithm}
\label{alg:binary_pprm}
\begin{algorithmic}[1]
\Require Input array $a$
\State blocksize $\gets 1$
\State $n \gets \log_2(\text{length of } a)$
\For{$k = 0$ to $n-1$}
    \State mask $\gets [[1] \times 2^k, [0] \times 2^k] \times 2^{n-(k+1)}$ \label{line:mask1}\\
\Comment{$mask:\underset{2^{k}}{\underbrace{111\ldots1}}\underset{2^{k}}{\underbrace{000\ldots0}}\:\cdots\:\underset{2^{k}}{\underbrace{111\ldots1}}\underset{2^{k}}{\underbrace{000\ldots0}}$} \label{line:mask2}
    \State temp $\gets \text{zeros like } a$ \Comment{Array of the same size as $a$}
    \State temp[blocksize:] $\gets (a \ \& \ \text{mask})[:-\text{blocksize}]$ \\
\Comment{Shift right masked $a$ by blocksize}
    \State $a \gets a \oplus \text{temp}$ \Comment{XOR operation between all blocks}
    \State blocksize $\gets 2 \times \text{blocksize}$
\EndFor
\Return $a$
\end{algorithmic}
\end{algorithm}

\section{Simulation Results and Analysis\protect\label{sec:Simulation}}

\begin{figure}[t]
\begin{centering}
\includegraphics[scale=0.6]{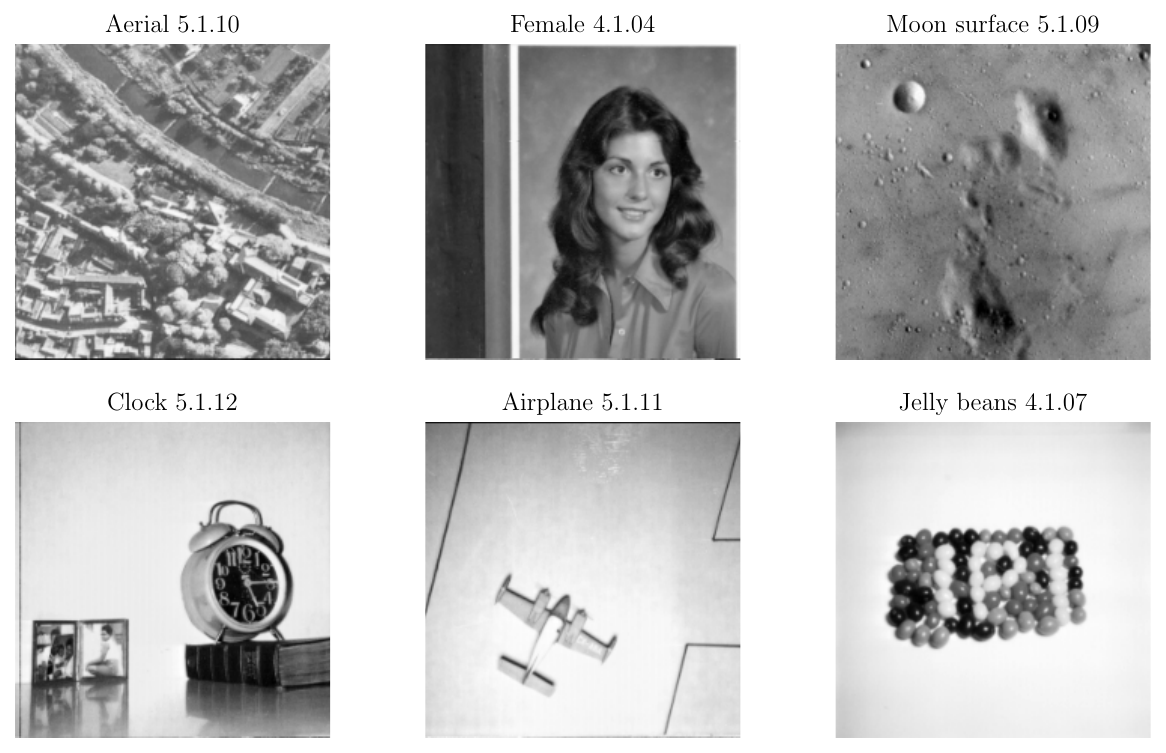}
\par\end{centering}
\caption{{\small Six tested images, each with a resolution of $256\times256$
\protect\label{fig:images}}}
\end{figure}

An experiment was conducted to test the PPRM QC optimization of the
NEQR representations of six images chosen from the USC-SIPI dataset
\cite{images}, each with a size of $256\times256$ and 256 grayscale
levels: \textquotedbl Aerial,\textquotedbl{} \textquotedbl Female,\textquotedbl{}
\textquotedbl Moon surface,\textquotedbl{} \textquotedbl Clock,\textquotedbl{}
\textquotedbl Airplane,\textquotedbl{} and \textquotedbl Jelly beans\textquotedbl{}
(Figure~\ref{fig:images}). To reduce the number of control qubits
in the MCNOT gates used in the NEQR circuit of each image, a $65536\times8$
(where $65536=256\times256$) binary matrix was prepared. Each row
of this matrix corresponds to a pixel and represents its 8-bit grayscale
value. Additionally, each column with index $i$ indicates which product
terms constitute the initial ESOP expression of $\ket{C_{i}^{YX}}$,
meaning that this column indicates the coefficients of the minterms
($a_{k}$'s in Equation~\ref{eq:minterms}) of the $f$ function corresponding
to the grayscale qubit $\ket{C_{i}^{YX}}$. 

Each column of this matrix is then transformed into a $\overrightarrow{b}$
vector using the \emph{Binary PPRM} function. Afterwards, the expected
PPRM expression for each of the eight grayscale qubits can be obtained
by applying the $*$ operation between $X(n)$ and each $\overrightarrow{b}$
vector. Then, the product terms in the $i$-th resulting PPRM expression
are used for the controlling qubits of the new MCNOT gates applied
to $\ket{C_{i}^{YX}}$.

Tables~\ref{tab:qc_mcnot} and~\ref{tab:qc_mcnotR} demonstrate the
reduction of QCs in each tested image using MCNOT and MCNOT-R gates,
respectively, along with their compression ratios, as defined in Equation~\ref{eq:compressionRate}.
On average, there is a 105.5-fold reduction in QC using MCNOT and
a 2.4-fold reduction in QC using MCNOT-R. The average compression
ratios are 99.05\% and 58.91\%, respectively.

\begin{equation}
Compression\,Ratio=\left(1-\frac{QC\,of\,Optimized}{QC\,of\,NonOptimized}\right)\times100\%\label{eq:compressionRate}
\end{equation}

\begin{table}[!t]
\begin{centering}
{\small{}%
\begin{tabular}{cccc}
\hline 
\multicolumn{1}{|c|}{{\small Image}} & \multicolumn{1}{c|}{\begin{cellvarwidth}[t]
\centering
{\small QC of}{\small\par}

{\small Non-Optimized}
\end{cellvarwidth}} & \multicolumn{1}{c|}{\begin{cellvarwidth}[t]
\centering
{\small QC of}{\small\par}

{\small Optimized}
\end{cellvarwidth}} & \multicolumn{1}{c|}{\begin{cellvarwidth}[t]
\centering
{\small Compression}{\small\par}

{\small Ratio}
\end{cellvarwidth}}\tabularnewline
\hline 
\hline 
{\small Aerial 5.1.10} & {\small 53865957128} & {\small 514219033 } & {\small 99.05\%}\tabularnewline
\hline 
{\small Female 4.1.04} & {\small 50562026908} & {\small 508945459} & {\small 98.99\%}\tabularnewline
\hline 
{\small Moon surface 5.1.09} & {\small 50633394160} & {\small 517464591} & {\small 98.98\%}\tabularnewline
\hline 
{\small Clock 5.1.12} & {\small 58309797340} & {\small 506508163} & {\small 99.13\%}\tabularnewline
\hline 
{\small Airplane 5.1.11} & {\small 56223632296} & {\small 516989043} & {\small 99.08\%}\tabularnewline
\hline 
{\small Jelly beans 4.1.07} & {\small 53324902920} & {\small 497631277} & {\small 99.07\%}\tabularnewline
\hline 
\end{tabular}}{\small\par}
\par\end{centering}
\caption{{\small PPRM QC Reductions of the NEQR Representations Using MCNOT
Gates \protect\label{tab:qc_mcnot}}}
\end{table}

\begin{table}[!t]
\begin{centering}
\begin{tabular}{cccc}
\hline 
\multicolumn{1}{|c|}{{\small Image}} & \multicolumn{1}{c|}{\begin{cellvarwidth}[t]
\centering
{\small QC of}{\small\par}

{\small Non-Optimized}
\end{cellvarwidth}} & \multicolumn{1}{c|}{\begin{cellvarwidth}[t]
\centering
{\small QC of}{\small\par}

{\small Optimized}
\end{cellvarwidth}} & \multicolumn{1}{c|}{\begin{cellvarwidth}[t]
\centering
{\small Compression}{\small\par}

{\small Ratio}
\end{cellvarwidth}}\tabularnewline
\hline 
\hline 
{\small Aerial 5.1.10} & {\small 74523104} & {\small 31476768} & {\small 57.76\%}\tabularnewline
\hline 
{\small Female 4.1.04} & {\small 69952144} & {\small 30483500} & {\small 56.42\%}\tabularnewline
\hline 
{\small Moon surface 5.1.09} & {\small 70050880} & {\small 31264133} & {\small 55.37\%}\tabularnewline
\hline 
{\small Clock 5.1.12} & {\small 80671120} & {\small 30194410} & {\small 62.57\%}\tabularnewline
\hline 
{\small Airplane 5.1.11} & {\small 77784928} & {\small 30924631} & {\small 60.24\%}\tabularnewline
\hline 
{\small Jelly beans 4.1.07} & {\small 73774560} & {\small 28723803} & {\small 61.07\%}\tabularnewline
\hline 
\end{tabular}
\par\end{centering}
\caption{{\small PPRM QC Reductions of the NEQR Representations Using MCNOT-R
Gates \protect\label{tab:qc_mcnotR}}}
\end{table}

\begin{figure}[!t]
\begin{centering}
\subfloat[]{\begin{centering}
\includegraphics[scale=0.55]{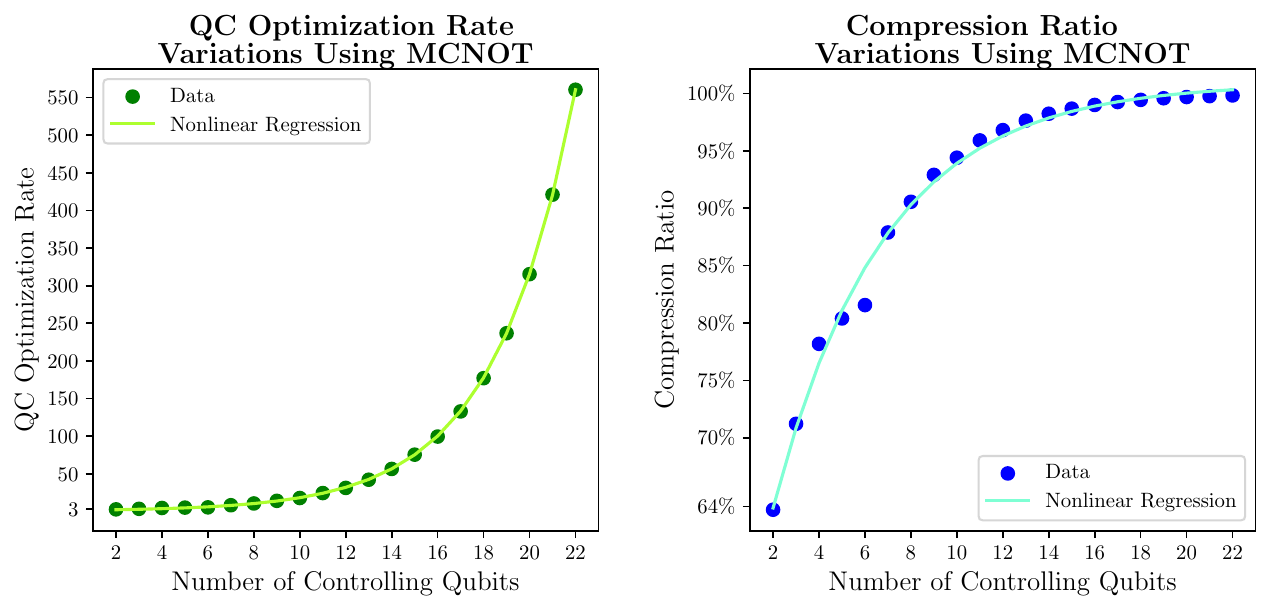}
\par\end{centering}
}\\\subfloat[]{\begin{centering}
\includegraphics[scale=0.55]{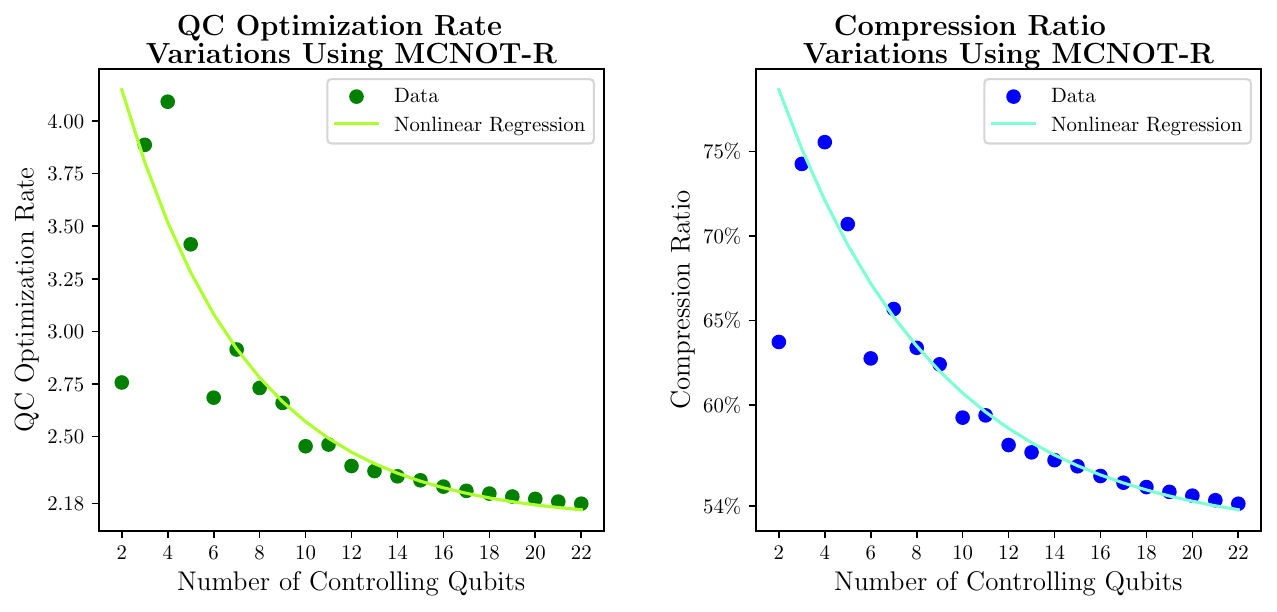}
\par\end{centering}

}
\par\end{centering}
\caption{{\small PPRM QC Optimization Rate and Compression Ratio for Random
Images sized from $2^{1}\times2^{1}$ to $2^{11}\times2^{11}$, using
(a) MCNOT gates and (b) MCNOT-R gates \protect\label{fig:opt}}}

\end{figure}

Another experiment investigates the QC optimization rates, defined
in Equation~\ref{eq:opt_rate}, and the compression ratios for images
with random grayscale pixel values, ranging in size from $2^{1}\times2^{1}$
to $2^{11}\times2^{11}$. These simulations were run on a system with
an Intel Core i3-6100 \texttimes{} 4 CPU, 8 GB RAM, and Linux Ubuntu
24.04.1 LTS, and images with dimensions beyond $2^{11}\times2^{11}$could
not be simulated due to hardware limitations. As shown in Figure~\ref{fig:opt},
considering $m$ as the number of controlling qubits $(2\leq m\leq22)$,
there is an exponential increase in the QC optimization rate for representations
using MCNOT gates, which is estimated by the function $1.33^{m}+0.49$.
In contrast, the optimization rate for representations using MCNOT-R
gates exhibits an exponential decrease, described by the function
$1.12^{-(1.60m-9.45)}+2.10$, with the curve approaching a limit of
2.10. In addition, the compression ratios for NEQR circuits using
MCNOT gates, estimated by the function $-2.43^{-(0.24m-4.54)}+100.89$,
asymptotically approach 100\% as the number of qubits increases. Similarly,
the compression ratios for NEQR circuits using MCNOT-R gates, estimated
by $1.82^{-(0.24m-5.93)}+52.27$, approach 52\% as the number of qubits
tends to infinity. These estimations were performed using the nonlinear
regression \emph{curve\_fit }method from the \emph{scipy.optimize
}library in Python. Consequently, the optimized QCs have complexities
of $O(1.5^{m})$ and $O(9m)$ for NEQR circuits using MCNOT and MCNOT-R
gates, respectively.

\begin{equation}
QC\,Optimization\,Rate=\frac{QC\,of\,NonOptimized}{QC\,of\,Optimized}\label{eq:opt_rate}
\end{equation}

\section{Conclusion \protect\label{sec:Conclusion}}

This paper aims to determine whether the PPRM transformation of ESOP
expressions in an NEQR quantum circuit can effectively reduce the
quantum cost (QC). To achieve this, an efficient algorithm in terms
of run-time and memory complexity was selected over another algorithm
that uses matrix multiplication. Six $256\times256$ test images were
transformed into binary vectors and converted to PPRM to verify their
QC reductions using either MCNOT gates or MCNOT-R gates in their NEQR
circuits. The results showed average QC reductions of 105.5 times
for NEQR circuits with MCNOT gates and 2.4 times for NEQR circuits
with MCNOT-R gates, with average compression ratios of 99.05\% and
58.91\%, respectively.

Additionally, images with random pixel values, with dimensions ranging
from $2^{1}\times2^{1}$ to $2^{11}\times2^{11}$, were analyzed to
investigate their QC reductions and compression ratios. The results
indicate that the optimization rate for NEQR circuits with exponential
QC $O(2^{m})$, which depends on the number of controlling qubits
$m$, increases exponentially and is estimated by $O(1.33^{m})$.
In contrast, the optimization rate for NEQR circuits with linear QC
$O(m)$ decreases as the number of controlling qubits increases, approaching
a constant value of $2.1$. Moreover, the compression ratios tend
to approach 100\% for circuits using MCNOT gates and 52\% for those
using MCNOT-R gates.


\begin{thebibliography}{36}
\providecommand{\natexlab}[1]{#1}
\providecommand{\url}[1]{\texttt{#1}}
\expandafter\ifx\csname urlstyle\endcsname\relax
  \providecommand{\doi}[1]{doi: #1}\else
  \providecommand{\doi}{doi: \begingroup \urlstyle{rm}\Url}\fi

\bibitem[Venegas-Andraca and Bose(2003)]{rep1}
Salvador Venegas-Andraca and Sougato Bose.
\newblock Storing, processing and retrieving an image using quantum mechanics.
\newblock \emph{Proceedings of SPIE - The International Society for Optical Engineering}, 5105, 08 2003.
\newblock \doi{10.1117/12.485960}.

\bibitem[Latorre(2005)]{rep2}
Jose~I. Latorre.
\newblock Image compression and entanglement, 2005.
\newblock URL \url{https://arxiv.org/abs/quant-ph/0510031}.

\bibitem[Venegas-Andraca and Ball(2010)]{rep8}
S.~E. Venegas-Andraca and J.~L. Ball.
\newblock Processing images in entangled quantum systems.
\newblock \emph{Quantum Information Processing}, 9\penalty0 (1):\penalty0 1--11, 2010.
\newblock ISSN 1573-1332.
\newblock \doi{10.1007/s11128-009-0123-z}.

\bibitem[Le et~al.(2011)Le, Dong, and Hirota]{rep3}
Phuc~Q. Le, Fangyan Dong, and Kaoru Hirota.
\newblock A flexible representation of quantum images for polynomial preparation, image compression, and processing operations.
\newblock \emph{Quantum Information Processing}, 10\penalty0 (1):\penalty0 63--84, 2011.
\newblock ISSN 1573-1332.
\newblock \doi{10.1007/s11128-010-0177-y}.

\bibitem[Sun et~al.(2013)Sun, Iliyasu, Yan, Dong, and Hirota]{rep9}
Bo~Sun, Abdullah Iliyasu, Fei Yan, Fangyan Dong, and Kaoru Hirota.
\newblock An {RGB} multi-channel representation for images on quantum computers.
\newblock \emph{Journal of Advanced Computational Intelligence and Intelligent Informatics}, 17:\penalty0 404--417, 05 2013.
\newblock \doi{10.20965/jaciii.2013.p0404}.

\bibitem[Zhang et~al.(2013)Zhang, Lu, Gao, and Wang]{neqr}
Yi~Zhang, Kai Lu, Yinghui Gao, and Mo~Wang.
\newblock {NEQR}: a novel enhanced quantum representation of digital images.
\newblock \emph{Quantum Information Processing}, 12\penalty0 (8):\penalty0 2833--2860, 2013.
\newblock ISSN 1573-1332.
\newblock \doi{10.1007/s11128-013-0567-z}.

\bibitem[Li and Liu(2018)]{rep6}
Panchi Li and Xiande Liu.
\newblock Color image representation model and its application based on an improved {FRQI}.
\newblock \emph{International Journal of Quantum Information}, 16\penalty0 (01):\penalty0 1850005, 2018.
\newblock \doi{10.1142/S0219749918500053}.

\bibitem[Jiang and Wang(2015)]{rep4}
Nan Jiang and Luo Wang.
\newblock Quantum image scaling using nearest neighbor interpolation.
\newblock \emph{Quantum Information Processing}, 14\penalty0 (5):\penalty0 1559--1571, 2015.
\newblock ISSN 1573-1332.
\newblock \doi{10.1007/s11128-014-0841-8}.

\bibitem[Jiang et~al.(2015)Jiang, Wang, and Mu]{rep7}
Nan Jiang, Jian Wang, and Yue Mu.
\newblock Quantum image scaling up based on nearest-neighbor interpolation with integer scaling ratio.
\newblock \emph{Quantum Information Processing}, 14, 08 2015.
\newblock \doi{10.1007/s11128-015-1099-5}.

\bibitem[Li et~al.(2019)Li, Fan, Xia, Peng, and Song]{rep5}
Hai-Sheng Li, Ping Fan, Hai-Ying Xia, Huiling Peng, and Shuxiang Song.
\newblock Quantum implementation circuits of quantum signal representation and type conversion.
\newblock \emph{IEEE Transactions on Circuits and Systems I: Regular Papers}, 66\penalty0 (1):\penalty0 341--354, 2019.
\newblock \doi{10.1109/TCSI.2018.2853655}.

\bibitem[Deb and Pan(2024)]{qic}
Sowmik~Kanti Deb and W.~David Pan.
\newblock Quantum image compression: Fundamentals, algorithms, and advances.
\newblock \emph{Computers}, 13\penalty0 (8), 2024.
\newblock \doi{10.3390/computers13080185}.
\newblock URL \url{https://www.mdpi.com/2073-431X/13/8/185}.

\bibitem[Pang et~al.(2006{\natexlab{a}})Pang, Zhou, and Guo]{compress4}
Chao Pang, Zheng-Wei Zhou, and Guang-Can Guo.
\newblock A hybrid quantum encoding algorithm of vector quantization for image compression.
\newblock \emph{Chinese Physics}, 15, 04 2006{\natexlab{a}}.
\newblock \doi{10.1088/1009-1963/15/12/044}.

\bibitem[Pang et~al.(2006{\natexlab{b}})Pang, Zhou, and Guo]{compress5}
Chao~Yang Pang, Zheng~Wei Zhou, and Guang~Can Guo.
\newblock Quantum discrete cosine transform for image compression, 2006{\natexlab{b}}.
\newblock URL \url{https://arxiv.org/abs/quant-ph/0601043}.

\bibitem[Nodehi et~al.(2009)Nodehi, Tayarani, and Mahmoudi]{compress6}
Ali Nodehi, Mohamad Tayarani, and Fariborz Mahmoudi.
\newblock A novel functional sized population quantum evolutionary algorithm for fractal image compression.
\newblock In \emph{2009 14th International CSI Computer Conference}, pages 564--569, 2009.
\newblock \doi{10.1109/CSICC.2009.5349639}.

\bibitem[Feng and Zhou(2014)]{compress7}
Qi-gao Feng and Hao-yu Zhou.
\newblock Research of image compression based on quantum bp network.
\newblock \emph{TELKOMNIKA Indonesian Journal of Electrical Engineering}, 12, 07 2014.
\newblock \doi{10.11591/telkomnika.v12i1.3908}.

\bibitem[Du et~al.(2015)Du, Yan, and Ma]{compress8}
Songlin Du, Yaping Yan, and Yide Ma.
\newblock Quantum-accelerated fractal image compression: An interdisciplinary approach.
\newblock \emph{IEEE Signal Processing Letters}, 22\penalty0 (4):\penalty0 499--503, 2015.
\newblock \doi{10.1109/LSP.2014.2363689}.

\bibitem[Jiang et~al.(2018)Jiang, Lu, Hu, Dang, and Cai]{compress1}
Nan Jiang, Xiaowei Lu, Hao Hu, Yijie Dang, and Yongquan Cai.
\newblock A novel quantum image compression method based on {JPEG}.
\newblock \emph{International Journal of Theoretical Physics}, 57\penalty0 (3):\penalty0 611--636, March 2018.
\newblock ISSN 1572-9575.
\newblock \doi{10.1007/s10773-017-3593-2}.

\bibitem[Ma and Zhou(2023)]{compress9}
Yan Ma and Nan-Run Zhou.
\newblock Quantum color image compression and encryption algorithm based on {F}ibonacci transform.
\newblock \emph{Quantum Information Processing}, 22\penalty0 (1):\penalty0 39, January 2023.
\newblock ISSN 1573-1332.
\newblock \doi{10.1007/s11128-022-03749-6}.

\bibitem[Wang et~al.(2024)Wang, Tan, Huang, and Zheng]{compress10}
Hengyan Wang, Jing Tan, Yixiao Huang, and Wenqiang Zheng.
\newblock Quantum image compression with autoencoders based on parameterized quantum circuits.
\newblock \emph{Quantum Information Processing}, 23\penalty0 (2):\penalty0 41, January 2024.
\newblock ISSN 1573-1332.
\newblock \doi{10.1007/s11128-023-04243-3}.

\bibitem[Ji et~al.(2024)Ji, Liu, Huang, Chen, and Wu]{compress11}
Xun Ji, Qin Liu, Shan Huang, Andi Chen, and Shengjun Wu.
\newblock Image compression and reconstruction based on quantum network, 2024.
\newblock URL \url{https://arxiv.org/abs/2404.11994}.

\bibitem[Nasr et~al.(2021)Nasr, Younes, and Elsayed]{compres2}
Norhan Nasr, Ahmed Younes, and Ashraf Elsayed.
\newblock Efficient representations of digital images on quantum computers.
\newblock \emph{Multimedia Tools and Applications}, 80, 10 2021.
\newblock \doi{10.1007/s11042-021-11355-4}.

\bibitem[Haque et~al.(2023)Haque, Paul, Ulhaq, and Debnath]{compres3}
Md~Ershadul Haque, Manoranjan Paul, Anwaar Ulhaq, and Tanmoy Debnath.
\newblock Advanced quantum image representation and compression using a dct-efrqi approach.
\newblock \emph{Scientific Reports}, 13, 03 2023.
\newblock \doi{10.1038/s41598-023-30575-2}.

\bibitem[Brayton et~al.(1984)Brayton, Hachtel, McMullen, and Sangiovanni-Vincentelli]{espresso}
Robert~K. Brayton, Gary~D. Hachtel, Curtis~T. McMullen, and Alberto~L. Sangiovanni-Vincentelli.
\newblock \emph{Logic Minimization Algorithms for {VLSI} Synthesis}.
\newblock Springer New York, NY, 1984.
\newblock \doi{10.1007/978-1-4613-2821-6}.

\bibitem[Khan(2019)]{rep10}
Rabia Khan.
\newblock An improved flexible representation of quantum images.
\newblock \emph{Quantum Information Processing}, 18, 05 2019.
\newblock \doi{10.1007/s11128-019-2306-6}.

\bibitem[Saini et~al.(2024)Saini, Behera, Al-Kuwari, and Farouk]{rep11}
Rakesh Saini, Bikash~K. Behera, Saif Al-Kuwari, and Ahmed Farouk.
\newblock {NEQRX}: Efficient quantum image encryption with reduced circuit complexity, 2024.
\newblock URL \url{https://arxiv.org/abs/2204.07996}.

\bibitem[Papakonstantinou and Papakonstantinou(2018)]{esop1}
Konstantinos Papakonstantinou and G.~Papakonstantinou.
\newblock A nonlinear integer programming approach for the minimization of {B}oolean expressions.
\newblock \emph{Journal of Circuits, Systems and Computers}, 27:\penalty0 1850163, 01 2018.
\newblock \doi{10.1142/S0218126618501633}.

\bibitem[Papakonstantinou(2014)]{esop2}
George Papakonstantinou.
\newblock A parallel algorithm for minimizing {ESOP} expressions.
\newblock \emph{Journal of Circuits, Systems and Computers}, 23, 02 2014.
\newblock \doi{10.1142/S0218126614500157}.

\bibitem[Becker et~al.(1994)Becker, Drechsler, and Theobald]{esop3}
Bernd Becker, Rolf Drechsler, and Michael Theobald.
\newblock Minimization of 2-level {AND/XOR} expressions using ordered {K}ronecker functional decision diagrams.
\newblock 05 1994.

\bibitem[Mishchenko and Perkowski(2001)]{esop4}
Alan Mishchenko and Marek Perkowski.
\newblock Fast heuristic minimization of {E}xclusive-{S}ums-of-{P}roducts.
\newblock \emph{5th International Reed-Muller Workshop}, 09 2001.

\bibitem[Lee et~al.(1999)Lee, Chung, Kim, and Lee]{karno}
Jae-Seung Lee, Yongwook Chung, Jaehyun Kim, and Soonchil Lee.
\newblock A practical method of constructing quantum combinational logic circuits, 1999.
\newblock URL \url{https://arxiv.org/abs/quant-ph/9911053}.

\bibitem[Iranmanesh et~al.(2022)Iranmanesh, Atta, and Ghanbari]{khodam}
Shahab Iranmanesh, Randa Atta, and Mohammad Ghanbari.
\newblock Implementation of a quantum image watermarking scheme using neqr on ibm quantum experience.
\newblock \emph{Quantum Information Processing}, 21\penalty0 (6):\penalty0 194, 2022.
\newblock ISSN 1573-1332.
\newblock \doi{10.1007/s11128-022-03530-9}.

\bibitem[Barenco et~al.(1995)Barenco, Bennett, Cleve, DiVincenzo, Margolus, Shor, Sleator, Smolin, and Weinfurter]{barenco}
Adriano Barenco, Charles~H. Bennett, Richard Cleve, David~P. DiVincenzo, Norman Margolus, Peter Shor, Tycho Sleator, John~A. Smolin, and Harald Weinfurter.
\newblock Elementary gates for quantum computation.
\newblock \emph{Physical Review A}, 52:\penalty0 3457--3467, 1995.
\newblock \doi{10.1103/PhysRevA.52.3457}.

\bibitem[Chrzanowska-jeske et~al.(2002)Chrzanowska-jeske, Mishchenko, and Perkowski]{family}
Malgorzata Chrzanowska-jeske, Alan Mishchenko, and Marek Perkowski.
\newblock Generalized inclusive forms; new canonical {R}eed-{M}uller forms including minimum {ESOP}s.
\newblock \emph{VLSI Design}, 14, 02 2002.
\newblock \doi{10.1080/10655140290009774}.

\bibitem[Khalid(2005)]{pprm_calc1}
Faraj Khalid.
\newblock \emph{Combinational logic synthesis based on the dual form of {R}eed-{M}uller representation}.
\newblock PhD thesis, Edinburgh Napier University, 2005.

\bibitem[Bakoev and Manev(2008)]{bakoev}
Valentin Bakoev and Krassimir Manev.
\newblock Fast computing of the positive polarity {R}eed-{M}uller transform over {GF} (2) and {GF} (3).
\newblock \emph{Eleventh International Workshop on Algebraic and Combinatorial Coding Theory}, pages 13--21, June 2008.

\bibitem[ima()]{images}
{USC-SIPI} {I}mage {D}atabase.
\newblock URL \url{https://sipi.usc.edu/database}.
\newblock Accessed on: Aug. 5, 2024.

\end{thebibliography}
\end{document}